%% file: main.tex
\documentclass[conference]{IEEEtran}
\usepackage{graphicx} 
\usepackage{subfig}
\usepackage{algorithm}
\usepackage{algorithmic}
\usepackage{hyperref}
\usepackage[round]{natbib}
\input{macros}

\let\cite\citep

\begin{document}

\title{Generalization Techniques Empirically Outperform\\ Differential Privacy against Membership Inference}
\author{\IEEEauthorblockN{Jiaxiang Liu}
\IEEEauthorblockA{University of Waterloo\\
 j632liu@uwaterloo.ca}
\and
\IEEEauthorblockN{Simon Oya}
\IEEEauthorblockA{University of Waterloo\\
simon.oya@uwaterloo.ca}
\and
\IEEEauthorblockN{Florian Kerschbaum}
\IEEEauthorblockA{University of Waterloo\\
florian.kerschbaum@uwaterloo.ca}}
\maketitle

\begin{abstract}
Differentially private training algorithms provide protection against one of the most popular attacks in machine learning: the membership inference attack.
However, these privacy algorithms incur a loss of the model's classification accuracy, therefore creating a privacy-utility trade-off.
The amount of noise that differential privacy requires to provide strong \emph{theoretical} protection guarantees in deep learning typically renders the models unusable, but authors have observed that even lower noise levels provide acceptable \emph{empirical} protection against existing membership inference attacks.

In this work, we look for alternatives to differential privacy towards empirically protecting against membership inference attacks.
We study the protection that simply following good machine learning practices (not designed with privacy in mind) offers against membership inference.
We evaluate the performance of state-of-the-art techniques, such as pre-training and sharpness-aware minimization, alone and with differentially private training algorithms, and find that, when using early stopping, the algorithms without differential privacy can provide both higher utility and higher privacy than their differentially private counterparts.
These findings challenge the belief that differential privacy is a good defense to protect against existing membership inference attacks.
\end{abstract}

\input{introduction}
\input{preliminaries}

\input{goal}
\input{setup}

\input{evaluation}
\input{relwork}
\input{conclusion}
\input{appendix}

\bibliographystyle{abbrvnat}
\bibliography{biblio}
\end{document}

%% file: macros.tex
\newcommand{\perror}{\text{P}_{\text{err}}}
\newcommand{\lrate}{\lambda}
\newcommand{\lreg}{\eta}
\newcommand{\Adv}{\textsc{Adv}}
\newcommand{\FPR}{\text{FPR}}
\newcommand{\FNR}{\text{FNR}}
\newcommand{\wei}{\mathbf{w}}
\newcommand{\loss}{L_\mathcal{S}}
\newcommand{\eee}{\mathbf{e}}

\usepackage{amsmath,amssymb,amstext, amsthm}
\newtheorem{defn}{Definition}[section]

%% file: introduction.tex
\section{Introduction}
Machine learning technologies are widely popular and we can find them as part of the services provided by major companies such as Amazon or Google.
It is well-known that machine learning models memorize information about the data they are trained with, and this makes them vulnerable to different attacks~\cite{song2017machine, carlini2019secret}.
One of the most prominent attacks against machine learning models is the Membership Inference Attack (MIA)~\cite{shokri2016membership, yeom2017privacy}, where an adversary that has access to a target model wants to determine whether or not a particular data sample was in the model's training set.
In many cases, leaking the membership of a sample can be highly sensitive (e.g., when the model has been trained with patients of a particular hospital).

Fortunately, the research community has developed defense mechanisms that protect against MIAs, such as adversarial regularization~\cite{nasr2018machine} and output obfuscation~\cite{jia2019memguard}.
Among existing defenses, those based on the notion of Differential Privacy (DP)~\cite{dwork2006calibrating} have been getting particular attention~\cite{chaudhuri2011differentially, abadi2016deep, yu2019differentially}, and popular open-source ML libraries like Tensorflow\footnote{\url{https://www.tensorflow.org/}} and PyTorch\footnote{\url{https://pytorch.org/}} already offer DP training algorithms.
DP randomizes the training process, statistically bounding the effect that a single data sample from the training set can have on the released model.
This naturally translates into protection against membership inference: given a model trained with differential privacy algorithms, it is hard for the adversary to tell whether or not a particular sample was in the training set.

One of the most efficient techniques to achieve DP in deep learning is the method proposed by \citet{abadi2016deep}, which provides a DP version of Stochastic Gradient Descent (SGD) known as DP-SGD.
This algorithm achieves its privacy guarantees by clipping the gradients and obfuscating them with Gaussian noise in each training epoch.
This noising operation also lowers the classification accuracy of the model, resulting in a \emph{privacy-utility trade-off}: increasing the variance of the Gaussian noise added in each iteration yields higher privacy levels (protection against MIA) but reduces the utility of the model (classification accuracy).

One of the main advantages of DP as a privacy notion is that it provides \emph{provable theoretical} privacy guarantees.
However, achieving such guarantees in deep learning requires noise levels that are so high that they render the model unusable for classification~\cite{jayaraman_evaluating_2019}, except for cases where the classification problem is very easy to learn~\cite{tramer2020differentially}.
However, researchers have shown that noise levels that keep the model's accuracy above reasonable levels, while not providing meaningful theoretical guarantees, can still successfully deter existing MIAs, thus achieving \emph{empirical privacy}~\cite{jayaraman_evaluating_2019}.
This suggests that DP and, in particular, DP-SGD, is a useful defense against existing MIAs.

In this work, we challenge this belief by evaluating DP-SGD along with other techniques that provide higher privacy and/or higher utility than DP-SGD.
We do not come up with new defenses, but instead evaluate existing techniques that are considered good practice in ML, such as pre-training, Sharpness-Aware Minimization (SAM), and early stopping.
Intuitively, these techniques contribute to \emph{generalizing} the model, i.e., they make the model perform well on samples it has not been trained on, while avoiding \emph{overfitting} the model to the training set.
These generalization techniques, that were designed to improve the model's classification accuracy (i.e., utility), also improve the protection against MIAs, since a generalized model will typically react similarly to members and non-members of the training set.
By increasing both privacy and utility, these techniques can break from the privacy-utility trade-off inherent to DP defenses.

We select two state-of-the-art MIAs and evaluate them against models trained with different generalization techniques and/or DP-SGD, studying how the privacy-utility trade-off varies after a certain number of training epochs.
We consider two datasets: CIFAR-100, an example of a complex classification problem, where previous works struggle to achieve both high privacy and utility~\cite{jayaraman_evaluating_2019}; as well as CIFAR-10, a simpler problem that allows high noise levels in DP-SGD without sacrificing a lot of utility.
In order to achieve reasonable classification accuracy levels in these datasets, we use models pre-trained on ImageNet,\footnote{\url{https://image-net.org/}} which increases utility while not affecting privacy.
We find that, in both datasets, DP is underwhelming in terms of the empirical protection it provides for a certain utility performance.
Among the different parameters of DP-SGD that we test, our experiments show that, surprisingly, setting the Gaussian noise level to zero provides the best privacy-utility trade-off.
Additionally, out of the generalization techniques we evaluate, we find that SAM performs best, outperforming DP-SGD.
We also study the effect of adding differentially private noise on top of the SAM optimizer, and find that the non-DP version still performs better.
We then study the performance of these algorithms in outliers, i.e., members and non-members that are particularly vulnerable to MIAs.
We find that, compared to our baseline, DP-SGD leaves outlier members more defenseless against MIAs, and provides higher protection to non-members instead.
This incurs serious privacy issues, since inferring membership is typically more privacy-sensitive than inferring non-membership.

In summary, our work questions the usefulness of DP techniques in deep learning to protect against MIAs: reasonable theoretical privacy guarantees require noise levels that destroy the model's accuracy, and empirical privacy performance can be achieved more efficiently by following good ML practices (not designed with privacy goals in mind).

%% file: preliminaries.tex
\section{Preliminaries}

In this section, we introduce our notation, explain membership inference attacks as well as differential privacy, and introduce the metrics that we use to assess the performance of machine learning models.

We use $z=(x,y)\in\mathcal{Z}$ to denote a data sample, which consists of a vector of features $x$ and a label $y$.
A training set $S$ is a set of data samples.
A training algorithm $A$ is a (possibly randomized) algorithm that receives a training set $S$ and outputs a trained model $a\in\mathcal{A}$.
A model $a$ evaluated on a vector of features $x$ returns a confidence score vector, with one confidence value for each class, which can be used to predict its label $y$.

\subsection{Membership Inference Attacks}
Membership Inference Attacks (MIAs) aim at guessing whether or not a particular data sample was part of the training set of a \emph{target model}.
Formally, given a data sample $z\in\mathcal{Z}$ and a trained model $a\in\mathcal{A}$, a MIA is a function $M(a,z)$ that outputs an estimation of whether or not $z$ was part of the training set of $a$.
We use $M(a,z)=1$ to denote that the MIA decided that $z$ was in the training set of $a$.
We assume the attack also knows the training algorithm $A$ and the training set size $n$~\cite{yeom2017privacy}, but we omit them from the arguments of $M$ for notational simplicity.

We consider only black-box MIAs, i.e., the attacker can query the model $a$ (any number of times) with a data sample to obtain confidence score vectors for that sample.
Black-box MIAs can broadly be classified into those that require to train Neural Networks (NN) to perform the inference, and those that do not.
\citet{shokri2016membership} proposed the first NN-based attack, also known as the Shadow Model attack.
This attack trains so-called shadow models on auxiliary data to mimic the behavior of the target model, and then trains an attacker model with the outputs of the shadow models, where the labels denote the membership attribute.
Finally, the attacker model is used on the outputs of the target model to predict the membership of samples.
\citet{salem2018ml} proposed a refinement of this attack that only requires one shadow model to function,  achieving similar performance.

Among the attacks that do not require training neural networks, the attack by \citet{yeom2017privacy} is perhaps the most popular one.
This attack considers that the adversary knows the loss function and the average loss of the training set.
The attack evaluates the target sample $z$ on the target model $a$, and decides that it is a member if its loss is below a threshold (typically, the average training set loss).
Despite its simplicity, this threshold attack achieves similar or even better performance than the NN-based attacks.
\citet{song2021systematic} recently showed that using different thresholds per class leads to a better performance.

All these attacks require access to the model's predicted confidence score, but recent works have proposed black-box attacks that only require access to the predicted labels~\cite{li2020membership,choquette2021label}.

In our evaluation, we consider two membership inference attacks that rely on the confidence score vector: the NN attack by \citet{salem2018ml} and the threshold attack by \citet{yeom2017privacy}.

\subsection{Privacy and Utility Metrics}
\label{sec:metrics}
We empirically assess the performance of a training algorithm $A$ by measuring the average privacy and utility of models $a$ produced with it.

\subsubsection{Privacy}
We measure the privacy of a model $a$ against an attack $M$ as the \emph{probability of error} of the attack against the model.
This error can be written in terms of two other metrics: the FPR and FNR.
The False Positive Rate (FPR) is the probability that the MIA decides that a non-member $z\notin S$ was a member of the training set, i.e., $\FPR\doteq\Pr(M(z,a)=1|z\notin S)$.
Likewise, the False Negative Rate (FNR) is the probability that the attack misclassifies a member of the training set as a non-member, i.e., $\FNR\doteq\Pr(M(z,a)=0|z\in S)$.
Finally, the probability of error is the average of these rates:
\begin{equation} \label{eq:perror}
    \perror\doteq\frac{\FPR+\FNR}{2}\,.
\end{equation}

This definition implicitly assumes a \emph{balanced} prior, i.e., this is the error when the true membership of the target sample $z$ is chosen by flipping an unbiased coin.
In Section~\ref{sec:out} we also study the protection that training algorithms offer over members and non-members separately.
Note that a naive adversary that randomly outputs a bit achieves $\perror=0.5$.
Therefore, $\perror$ typically ranges from 0 (no privacy) to 0.5 (total privacy).
We note that other works~\cite{yeom2017privacy, jayaraman_evaluating_2019} use a \emph{privacy loss} metric known as the \emph{membership advantage} ($\Adv$), which can be written as $\Adv=1-2\perror$.
Our findings also extend to this metric, since $\perror$ and $\Adv$ are linearly related.

\subsubsection{Utility}
We measure the utility of a model $a$ as the probability that the model correctly classifies a data sample that was not in the training set, also known as the \emph{testing accuracy}.
This utility metric measures how useful the model is towards the classification task for which it has been trained, and ranges from 0 (no utility) to 1 (perfect utility).

\subsubsection{Algorithm Comparison}
Given two training algorithms $A$ and $A'$, we say that $A$ outperforms $A'$ if it achieves higher privacy and no less utility than $A'$, or higher utility and no less privacy than $A'$.

\subsection{Differential Privacy in Machine Learning}
We define DP in the context of membership inference attacks.
Let $S\in\mathcal{Z}^{n}$ be a dataset of $n$ samples.
We say $S$ and $S'$ are \emph{neighbouring} datasets if $S'$ is a copy of $S$ with one sample added or removed.
Differentially private training algorithms $A$ ensure that neighbouring datasets have similar probabilities of producing a model $a\in\mathcal{A}$.
Formally,
\begin{defn}[($\epsilon,\delta$)-DP]
\label{defn:dp}
    A training algorithm $A$ is ($\epsilon$,$\delta$)-DP if, for every pair of neighbouring datasets $S$ and $S'$, and any subset of trained models $\mathcal{R}\subseteq{A}$, the following condition holds:
    \begin{equation} \label{eq:DP}
        \Pr(A(S)\in\mathcal{R})\leq \Pr(A(S')\in\mathcal{R})\cdot e^\epsilon + \delta\,.
    \end{equation}
\end{defn}
The parameter $\epsilon$ characterizes the privacy level, and $\delta$ is a slack variable that relaxes the condition and should be chosen to be smaller than $1/n$~\cite{dwork2014algorithmic}.
Small values of $\epsilon$ force the probabilities in \eqref{eq:DP} to be closer, and therefore denote high privacy regimes, while large values of $\epsilon$ denote low privacy regimes.

DP is a strong worst-case protection in the sense that, even if the adversary somehow knows $S$ and $S'$, deciding which of the two is the actual training set after observing the model $a$ is non-trivial.

One of the most efficient approaches to provide DP during the training process~\cite{chaudhuri2011differentially, al2019privacy} is the DP version of Stochastic Gradient Descent (SGD) by \citet{abadi2016deep}.
This algorithm, known as DP-SGD, follows the standard SGD algorithm, working in steps or batches.
In each batch, the algorithm selects a subset of the training samples and computes the gradient of the loss function evaluated on each.
Then, instead of averaging these gradients to update the model's weights (like standard SGD), DP-SGD first clips the norm of the gradients using a threshold $C$, averages the clipped gradients, applies Gaussian noise with standard deviation $C\cdot\sigma$ to them, and then updates the weights.
Here, $\sigma$ is known as the \emph{noise multiplier} and tunes the amount of noise that the algorithm adds.
The clipping plus noising step ensures DP guarantees.
Following the moments accountant method \cite{abadi2016deep}, one can compute the actual $(\epsilon,\delta)$ guarantee of DP-SGD, which depends on $\sigma$ as well as other model parameters like the training set size and number of epochs.

%% file: goal.tex
\section{Problem Statement}
In this section, we explain the limitations of DP as a theoretical protection against MIA, and argue that existing generalization techniques might empirically perform better than DP.
This motivates our experiments below.

\subsection{Limitations of Differential Privacy}

The main strength of DP, that sets it apart from other privacy notions, is that it provides \emph{provable theoretical guarantees} against attacks.
\citet{kairouz2017composition} show that an $(\epsilon,\delta)$-DP algorithm ensures the following  bounds involving the FNR and FPR of an adversary trying to distinguish between the neighbouring datasets, which by extension applies to MIAs:
\begin{equation}\label{eq:bounds}
    \FPR+e^\epsilon\cdot \FNR \geq 1 - \delta\,,\qquad
    \FNR+e^\epsilon\cdot \FPR \geq 1 - \delta\,.
\end{equation}
Adding these two expressions together and writing the bound in terms of the probability of error $\perror$ in \eqref{eq:perror}, we get:
\begin{equation} \label{eq:bound}
    \perror \geq \frac{1-\delta}{1+e^\epsilon}\,.
\end{equation}
Recall that the probability of error of an adversary that runs an attack no worse than random guessing is $0\leq\perror\leq 0.5$.
Assuming a very small $\delta\ll 1$, the bound in \eqref{eq:bound} means that a small $\epsilon=0.1$ ensures that the adversary's error will be very close to random guessing $\perror\geq 0.47$, while an $\epsilon=1$ provides $\perror\geq 0.26$.
A higher value like $\epsilon=5$ can only ensure $\perror\geq0.0067$, which is certainly a very weak guarantee, close to the trivial bound $\perror\geq 0$.
We can reasonably conclude that the theoretical protection that DP offers for values of $\epsilon>1$ is not strong, and stops being useful altogether when $\epsilon>5$.
Values of $\epsilon<5$ can only be achieved while providing reasonable test accuracy levels in simple classification tasks, such as MNIST and CIFAR-10.
More complex tasks like CIFAR-100 require higher $\epsilon$ values~\cite{jayaraman_evaluating_2019}, which cannot provide useful theoretical protection guarantees.

The actual \emph{empirical} privacy that DP algorithms provide against \emph{existing} MIAs, however, is significantly above the bound in \eqref{eq:bound}.
This is not because the bound is loose (\citet{nasr2021adversary} proved the bounds in \eqref{eq:bounds}, and by extension \eqref{eq:bound}, are indeed tight), but because the bound holds against a rather unrealistic adversary that knows all the entries in the dataset but one (i.e., $S$ and $S'$ in our notation).
In their evaluation, \citet{jayaraman_evaluating_2019} show that even values of $\epsilon=100$ or even $\epsilon=1\,000$ provide a significant amount of protection against existing (realistic) MIAs, while keeping utility high.
This suggests that DP is a reasonable empirical protection guarantee, if the data owner can afford its privacy-utility trade-off.

\subsection{Generalization Techniques as an Alternative to Differential Privacy in Deep Learning}
Previous work shows that a model's vulnerability against MIAs is connected to overfitting~\cite{yeom2017privacy}, and it is well-known that generalization techniques can help against MIAs~\cite{choquette2021label, li2020membership, nasr2018machine, salem2018ml}.
\citet{song2021systematic} recently showed that early stopping alone can outperform ad-hoc defenses like adversarial regularization~\cite{nasr2018machine}.

Motivated by this, in this work we evaluate the privacy and utility guarantees of existing off-the-shelf generalization techniques, and compare them with differentially private training.
There is a plethora of ML techniques and model configurations that provide high generalization and utility performance.
We pick a small subset of these techniques only to illustrate why there are alternatives better than DP when aiming for empirical privacy, and leave a more detailed evaluation as subject for future work.
A key aspect of our evaluation is that we assess the privacy-utility trade-off of our models \emph{after every training epoch}, so that we better understand the effect that early stopping has in the model's performance.

%% file: setup.tex
\section{Experimental Setup}

Our code runs in Python and we use PyTorch ML library\footnote{\url{https://pytorch.org/}}.
We use the DP-SGD implementation from Opacus DP library\footnote{https://opacus.ai/}, version 0.14, which follows the algorithm by \citet{abadi2016deep}.
We ran our experiments in a private computer server and Google Colab, both of which use Tesla P100 GPUs.
(We note that running times are not relevant for this paper.)
Our code will be available upon request.

\paragraph{Datasets}
We consider two datasets for our evaluation: CIFAR-100 and CIFAR-10.\footnote{\url{https://www.cs.toronto.edu/~kriz/cifar.html}}
CIFAR-100 contains $60\,000$ $32\times 32$ RGB images distributed into 100 different classes.
The dataset is already divided into a (default) training set of $50\,000$ images, and a (default) testing set of $10\,000$ images; each set has even representation of each class.
CIFAR-10 is similar to CIFAR-100, but with 10 classes instead.

For both datasets, we split the default training and testing set in half.
We follow a stratified split, i.e., we run the split \emph{per class}, ensuring every class has the same representation in our halves.
We use one half of the default training set as our actual training set (i.e., the members), and one half of the default testing set as our testing set (which are also the non-members).
We use the remaining halves of the built-in training and testing sets to train the neural-network MIA by~\citet{salem2018ml}.


\paragraph{Membership Inference Attacks}
We consider two state-of-the-art MIAs: the \emph{Neural Network (NN) attack} by \citet{salem2018ml} and the \emph{threshold attack} by \citet{yeom2017privacy}.
In order to measure privacy, we measure the FPR of the attack by evaluating it on the non-members, and the FNR by evaluating it on the members.
Then, we compute the probability of error $\perror$ from the FPR and FNR following \eqref{eq:perror}.

\paragraph{Model configuration}
Our model configuration choices are based on state-of-the-art results for CIFAR-100 by~\citet{foret2020sharpness}.
Even though \citeauthor{foret2020sharpness} achieve the best results using an EffNet model, this is not directly compatible with Opacus' DP-SGD implementation.
Instead, we use a ResNet50 architecture pre-trained on ImageNet.\footnote{\url{https://github.com/ppwwyyxx/GroupNorm-reproduce/releases/tag/v0.1}} 
This model is pre-trained using group normalization~\cite{wu2018group} instead of batch normalization, since the latter is not compatible with Opacus.
\citeauthor{wu2018group} show group and batch normalization achieve similar performances on ImageNet.


Our baseline experiments use the SGD optimizer, and we also test DP-SGD from the Opacus library and the Sharpness-Aware Minimization (SAM) optimizer~\cite{foret2020sharpness}.
SAM modifies SGD by adding an extra regularization term to the loss function that makes the algorithm look for a minimum whose neighbourhood is also at a low loss value (i.e., a ``wide'' minimum).
We use a PyTorch implementation of SAM,\footnote{\url{https://github.com/davda54/sam}} and set its hyperparameter (neighbourhood size) to $\rho=0.05$, following \citet{foret2020sharpness}.
We also use an $\ell_2$ regularization coefficient of $\lreg=0.0005$ by default in all of our experiments (unless mentioned otherwise).
We train each model for 15 epochs, using the step decay for the learning rate from the SAM implementation: we start with a rate of $\lrate$ that we keep for the first five epochs, the next four epochs use $0.2\cdot\lrate$, the next three use $0.2^2\cdot\lrate$, and the last three $0.2^3\cdot\lrate$.
We use a batch size of 32.


Besides SAM, we consider two generalization techniques also evaluated in previous work: $\ell_2$ regularization~\cite{krogh1992simple} and dropouts~\cite{srivastava2014dropout}.

We save the model after each epoch in order to understand how the privacy and utility guarantees of the model evolve with training.
We repeat all of our experiments 10 times to account for the randomness in training, and plot average results as well as $95\%$ confidence intervals.

%% file: evaluation.tex
\section{Evaluation}

\begin{figure}
    \centering
    \includegraphics[width=0.8\linewidth]{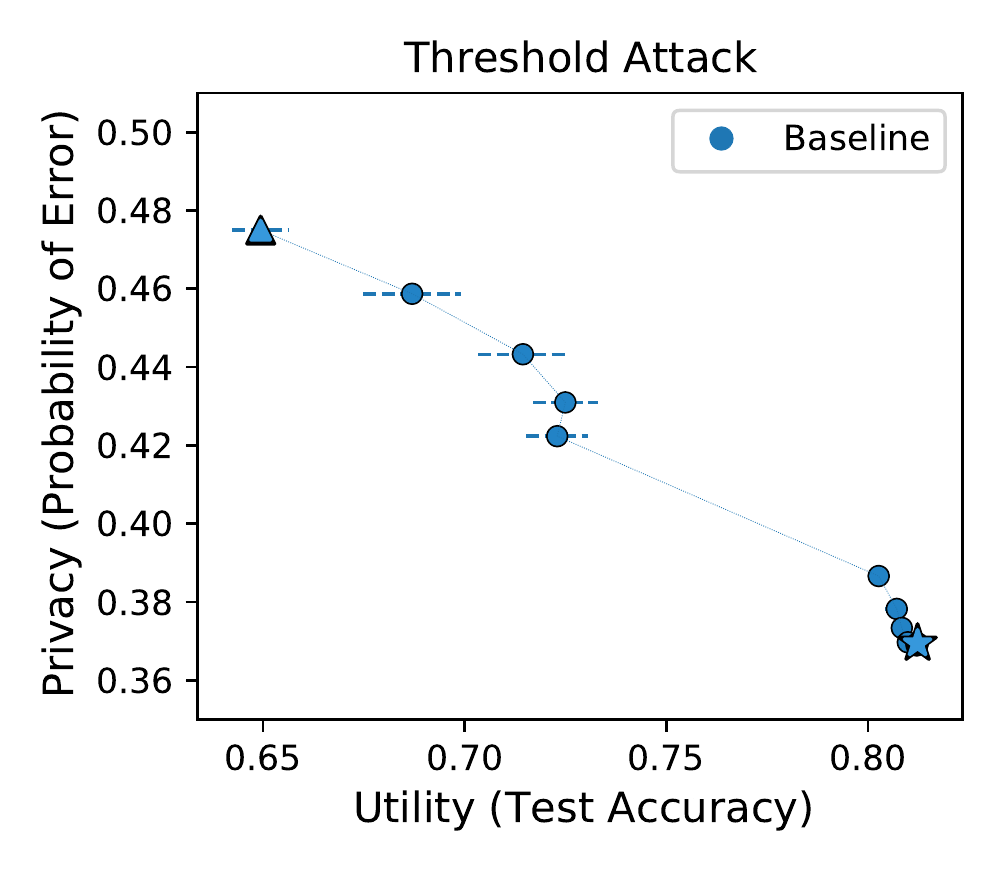}
    \caption{Baseline experiment.}
    \label{fig:exp0_yeom}
\end{figure}

We first consider the CIFAR-100 dataset, which is a more realistic and harder classification problem than CIFAR-10.
We begin our evaluation with a simple experiment to explain our plotting methodology.
Figure~\ref{fig:exp0_yeom} shows the performance of our baseline experiment (SGD optimizer) with an initial learning rate of $\lrate=0.01$  against the threshold attack.
Each point in the plot represents the performance (privacy vs.~utility) of the model after one epoch (a total of 15 points).
We represent the results of the first epoch with a triangle ($\triangle$) and the last epoch with a star ($\star$).
Intermediate epochs ($\bullet$) are joined by a thin dotted line.
The crosses behind the points show the $95\%$ confidence intervals for the privacy (vertical line) and the utility (horizontal line).
Note that when these intervals are not visible, it means they are smaller than the marker size.

As the number of epochs increases, the model memorizes more about the training set.
This causes an increase in utility and a decrease in privacy.
The large gap between the fifth and sixth epochs is due to the change in learning rate that happens at that point; we see this effect in our other experiments as well.
We see that, after 15 epochs, the model reaches an accuracy of $81\%$, which is a high value for this dataset (note that our training set is only of size $25\,000$, since we do not use all $50\,000$ samples for training).

We ran all of our experiments for the threshold and NN attacks, but found only small differences between them.
Due to space constraints, we show only the performance of the threshold attack in this paper, with the exception of Figure~\ref{fig:exp31}, where we also include the NN attack to illustrate the similarities in their performance.
We include a side-to-side comparison between the threshold and NN attacks in all our experiments in Appendix~\ref{app:results}.

\subsection{Effect of Differential Privacy}

We perform an initial experiment to look for reasonable values for the learning rate $\lrate$ and the clipping threshold $C$ for DP-SGD in CIFAR-100 (Figure~\ref{fig:exp21_yeom}).
We fix the noise multiplier $\sigma=0.01$ and $\delta=10^{-5}$.
This $\sigma$ is low enough so that the test accuracy does not decrease significantly, and we empirically see it provides high privacy levels $\perror>0.4$.
The $\delta$ ensures that $\delta<1/n=4\cdot10^{-5}$, which is recommended~\cite{dwork2014algorithmic}.
For reference, this configuration yields an $\epsilon\approx186\,000$ after 15 epochs, which does not provide a meaningful theoretical protection.
Figure~\ref{fig:exp21_yeom} shows that, out of the $\lrate$ and $C$ values we tried (in the legend), $C=1$ and $\lrate=0.01$ (green markers) achieves the best results: for every non-green point in the plot, there is a green point that achieves (roughly) higher privacy and utility.
We verified the same is true when setting $\sigma=0.02$.
Therefore, we fix $\lrate=0.01$ and $C=1$ for our remaining experiments in CIFAR-100.

Next, we evaluate the effect of different noise multipliers $\sigma$ in DP-SGD.
We set $C=1$, $\lrate=0.01$, and $\delta=10^{-5}$, and vary $\sigma\in\{0, 0.01, 0.02, \dots, 0.05\}$.
We show the results in Figure~\ref{fig:exp22_yeom}.
As expected, increasing $\sigma$ improves the privacy protection and decreases the utility of the model.
We can see this, for example, by looking at the final trained model ($\star$).
However, the privacy-utility trade-off caused by increasing $\sigma$ is worse than that caused by early stopping.
Roughly speaking, for every point in the graph, there is a $\sigma=0$ point that achieves no less privacy and utility than it.
This means that, paradoxically, the best DP-SGD configuration is the one that does not provide differential privacy at all ($\sigma=0$ implies $\epsilon=\infty$).

This experiment also confirms that it is not possible to achieve theoretical protection ($\epsilon<5$) in CIFAR-100 while keeping a reasonable utility loss: even the largest $\sigma$ we test ($\sigma=0.05$) yields a huge $\epsilon\approx 7\,000$.

\subsection{Comparison of Regularization Techniques}
\begin{figure*}
	\centering
	\begin{minipage}{0.32\linewidth}
	\centering
		\includegraphics[width=\linewidth]{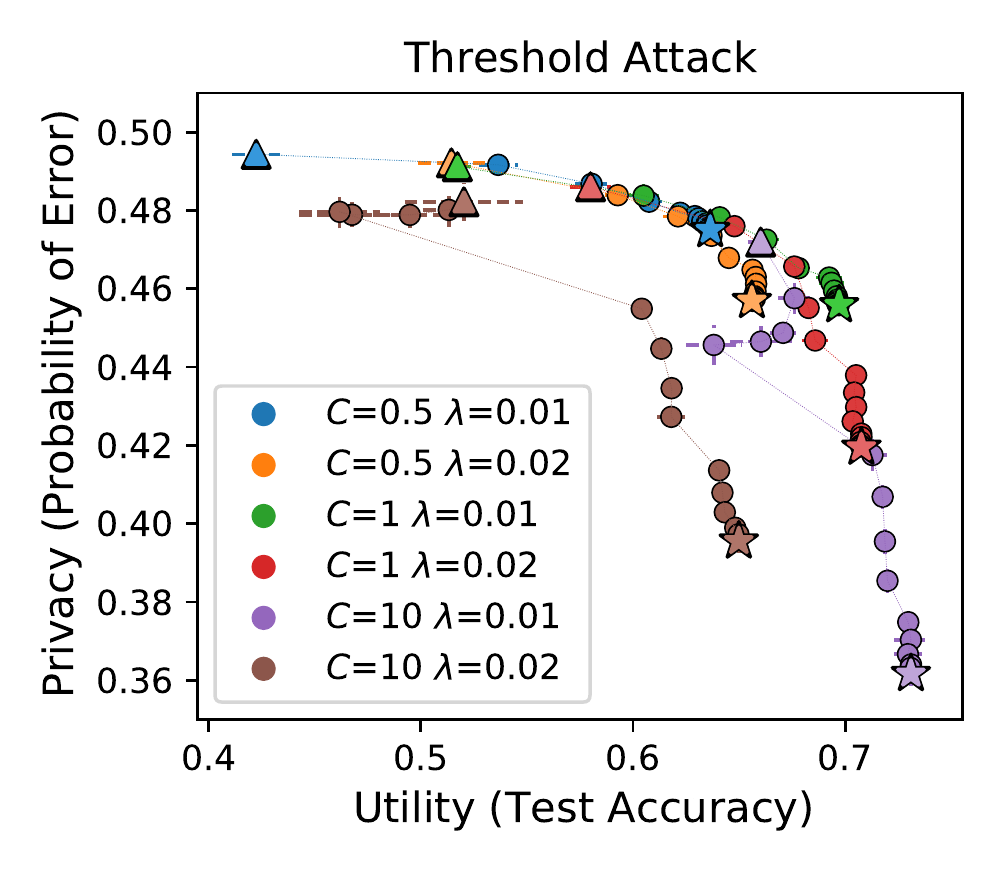}
		\caption{DP-SGD with different clipping thresholds $C$ and learning rates $\lrate$ (noise magnitude $\sigma=0.01$).}\label{fig:exp21_yeom}
	\end{minipage} \hfill
	\begin{minipage}{0.32\linewidth}
	\centering
		\includegraphics[width=\linewidth]{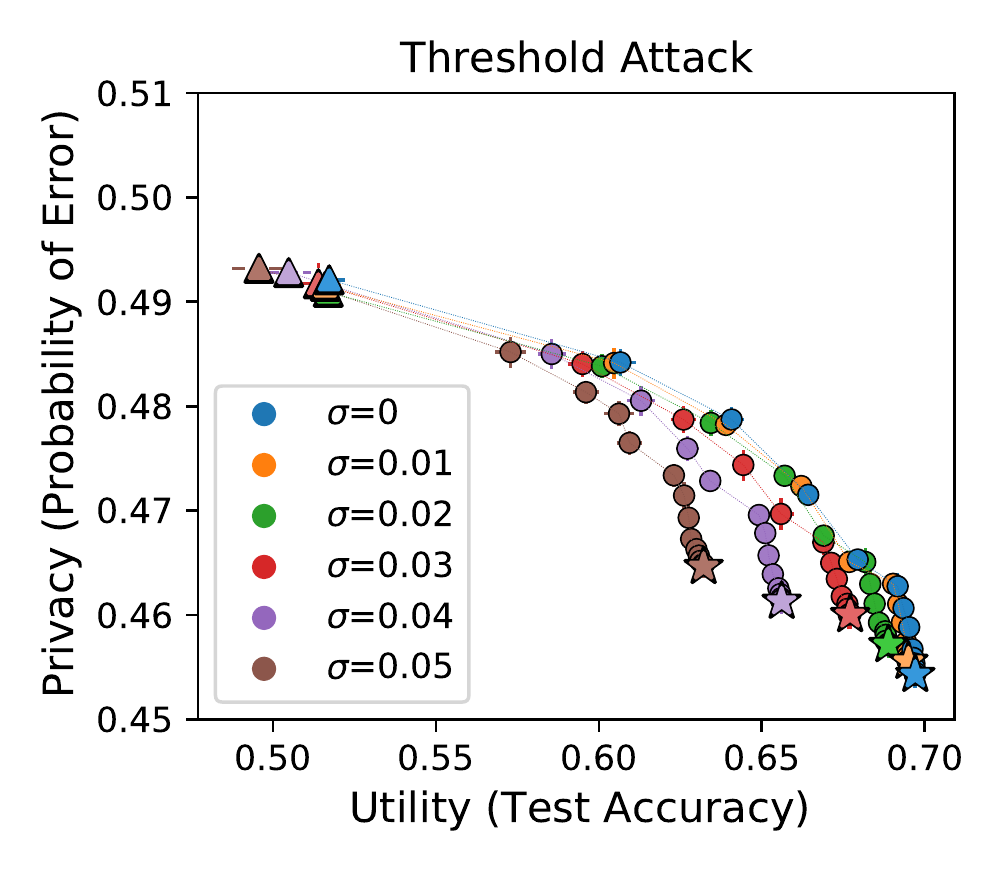}
		\caption{DP-SGD with different noise magnitudes $\sigma$ ($C=1$, $\lrate=0.01$).\\}\label{fig:exp22_yeom}
	\end{minipage}
	\hfill
	\begin{minipage}{0.32\linewidth}
	    \centering
		\includegraphics[width=\linewidth]{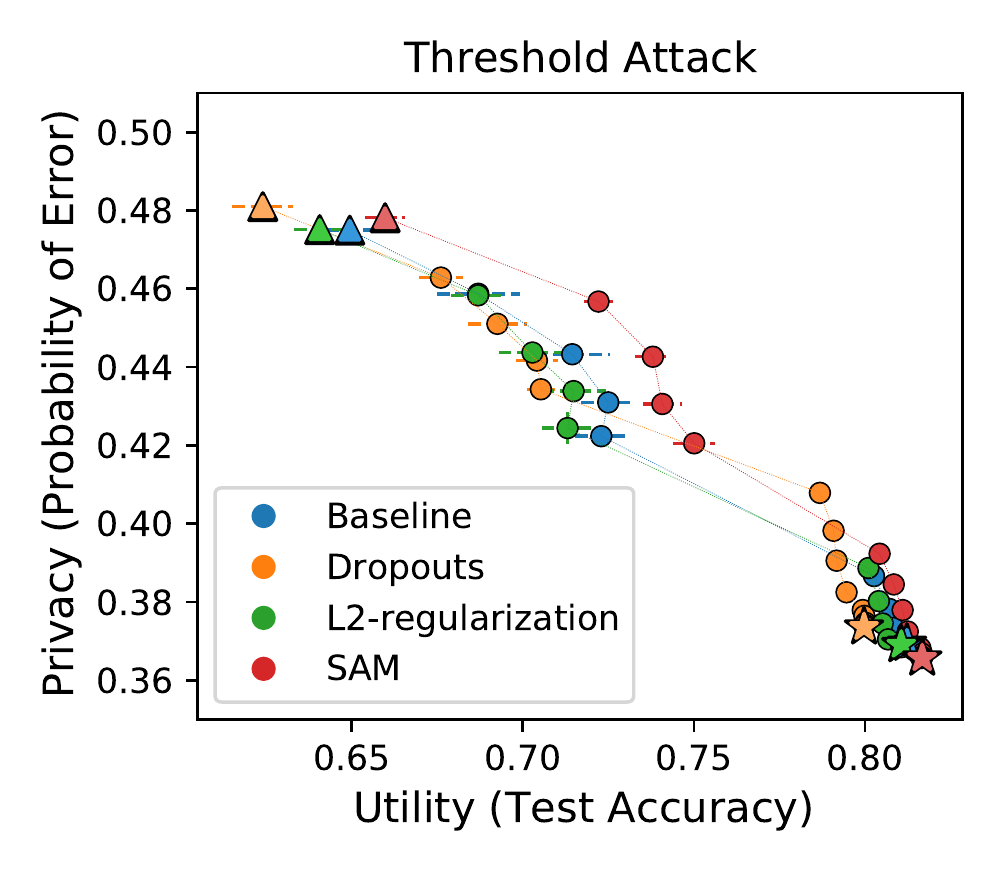}
		\caption{Privacy-utility trade-off for different regularization techniques.\\}\label{fig:exp1_yeom}
	\end{minipage}
\end{figure*}

Next, we compare between three regularization techniques:
\begin{itemize}
    \item $\ell_2$-regularization: for this experiment, we use SGD with an $\ell_2$ regularization coefficient of $\lreg=0.025$.
    \item Dropouts: we add a dropout layer with dropout rate of 0.6 before the last layer of the pre-trained ResNet50 model, and train using the SGD optimizer.
    \item SAM: we use the Sharpness-Aware Minimization optimizer with hyperparameter $\rho=0.05$.
\end{itemize}
Figure~\ref{fig:exp1_yeom} shows these results, as well as the baseline SGD.
If we were to judge by the last epoch only ($\star$), we see that SAM leaks the most, followed by the baseline and $\ell_2$-regularization, and then the dropouts.
However, using the last epoch as comparison is misleading: by looking at the evolution of the privacy-utility trade-off with the training epochs, we see that SAM provides the best performance.

\subsection{Differential Privacy vs.~Regularization}
\begin{figure*}[t]
	\centering
	\hfill
	\begin{minipage}{0.64\linewidth}
    	\begin{minipage}{0.48\linewidth}
    	    \centering
    		\includegraphics[width=\linewidth]{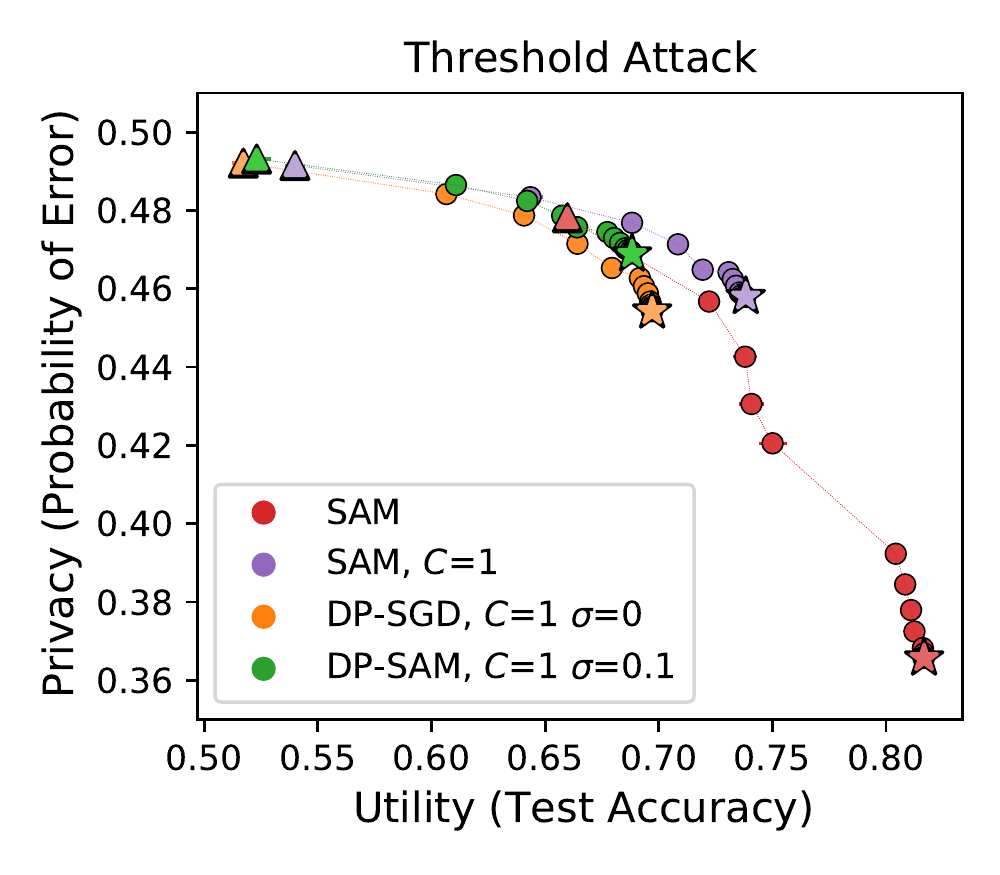}\\
    (a) Threshold Attack
    	\end{minipage}\hfill
    	\begin{minipage}{0.48\linewidth}
    	    \centering
    		\includegraphics[width=\linewidth]{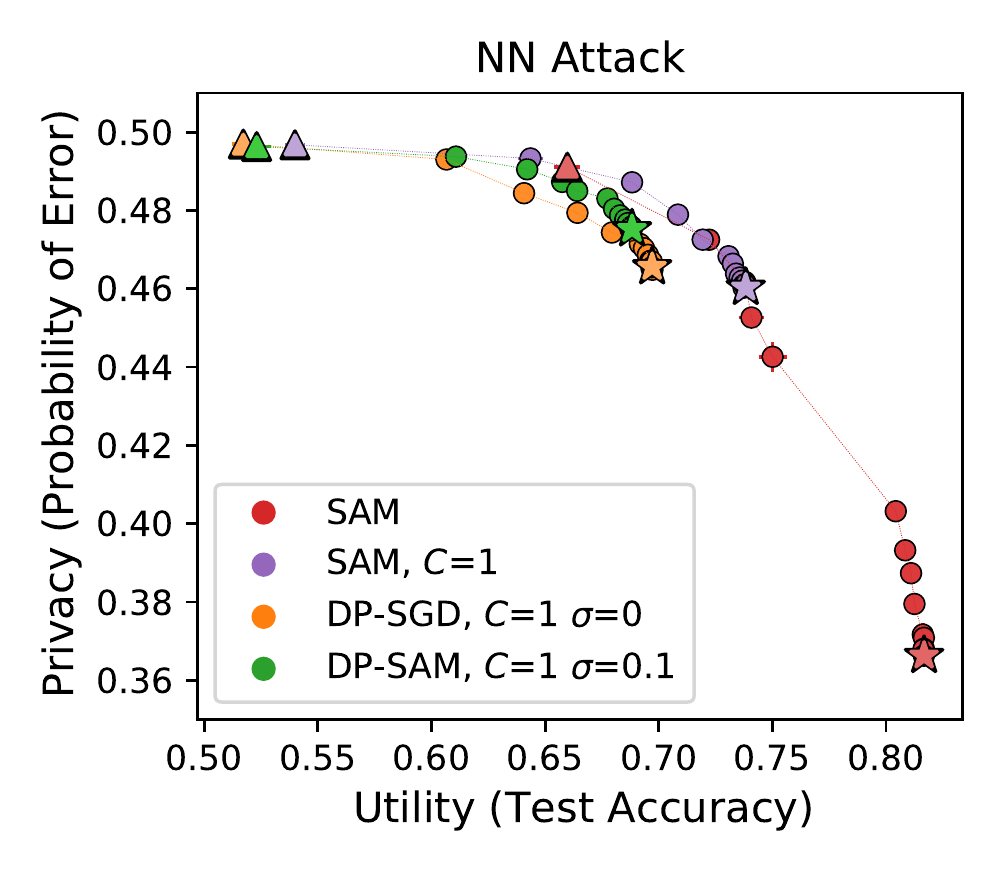}\\
    		(b) NN Attack
    	\end{minipage}
    	\caption{Comparison between SAM, SAM with clipping, DP-SGD, and DP-SAM in CIFAR-100.}\label{fig:exp31}
	\end{minipage} \hfill
	\begin{minipage}{0.32\linewidth}
	    \centering
		\includegraphics[width=\linewidth]{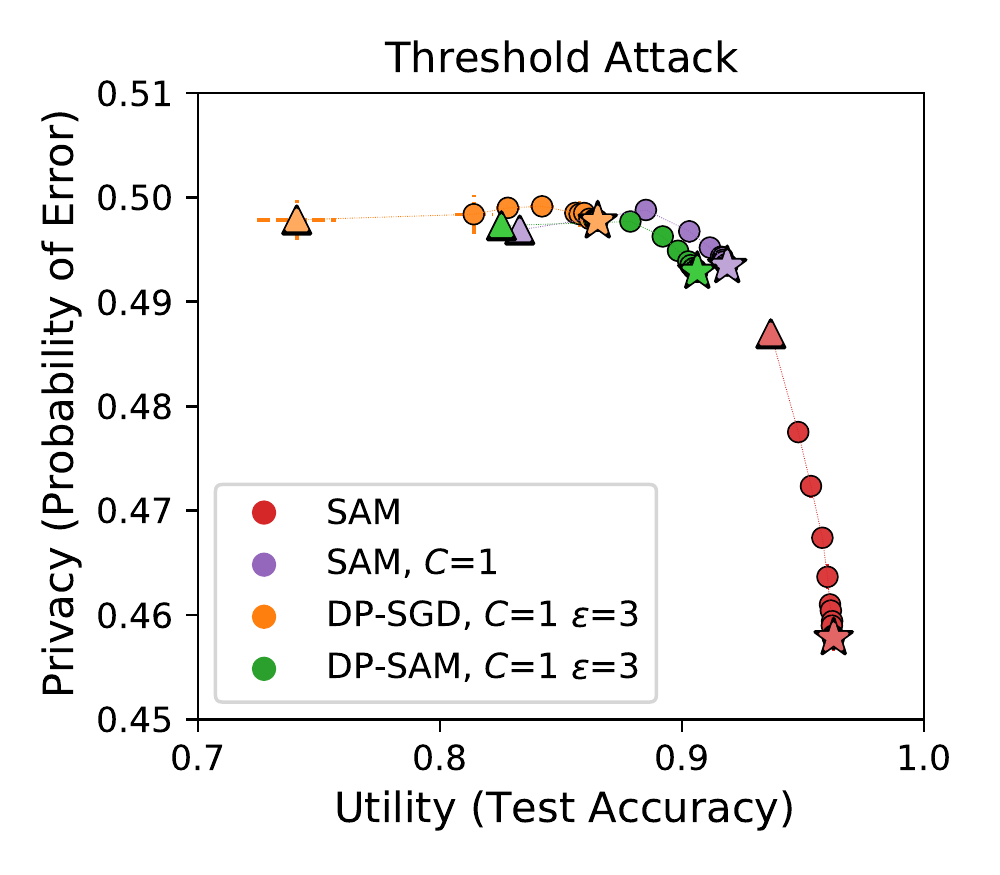}
		\caption{Comparison between SAM, SAM with clipping, DP-SGD, and DP-SAM in CIFAR-10.}\label{fig:exp42_yeom}
	\end{minipage}
	\hfill
\end{figure*}

Next, we compare the best-performing DP-SGD model (i.e., $\sigma=0$) with the best-performing regularization technique that we have tried (i.e., SAM).
We show the results in Figure~\ref{fig:exp31} for both the threshold and NN attacks.
We cannot compare DP-SGD and SAM directly, since the models produced with SAM are in a lower privacy and higher utility region than the models produced with DP-SGD.
This is because, with our parameters, SAM learns too fast.
In order to compare them, we found that clipping the gradients of SAM provides utility-privacy points in the same region as DP-SGD.
When comparing SAM with $C=1$ and DP-SGD, we see that SAM clearly outperforms the latter.

To further confirm that adding noise to the gradients is not a good idea, we ran a differentially private version of SAM that we call DP-SAM.
We achieve this in our code by simply plugging Opacus's Privacy Engine into the SAM optimizer (for completeness, we include the pseudo-code of this algorithm in Appendix~\ref{app:DP-SAM}).
We use a noise multiplier of $\sigma=0.1$, and we see that the effect of adding noise is the same that we saw in Figure~\ref{fig:exp22_yeom}: given a single epoch, DP-SAM achieves more privacy and less utility than SAM with clipping, but when considering early stopping, SAM with clipping $C=1$ clearly outperforms DP-SAM.

\subsection{Outliers Protection}
\label{sec:out}

Previous works show that, even when the model generalizes well on average, there are certain samples, also called \emph{outliers}, that can still be consistently unprotected~\cite{long2018understanding,choquette2021label}.
Measuring MIA performance with average metrics such as $\perror$ would not show such vulnerable samples.
In this section we identify outliers in our previous experiments, and study whether DP can protect such samples.
We consider that a sample is an outlier if it was correctly identified by our attack across all 10 runs.

\begin{figure*}[!t]
	\centering
	\begin{minipage}{0.32\linewidth}
	\centering
		\includegraphics[width=\linewidth]{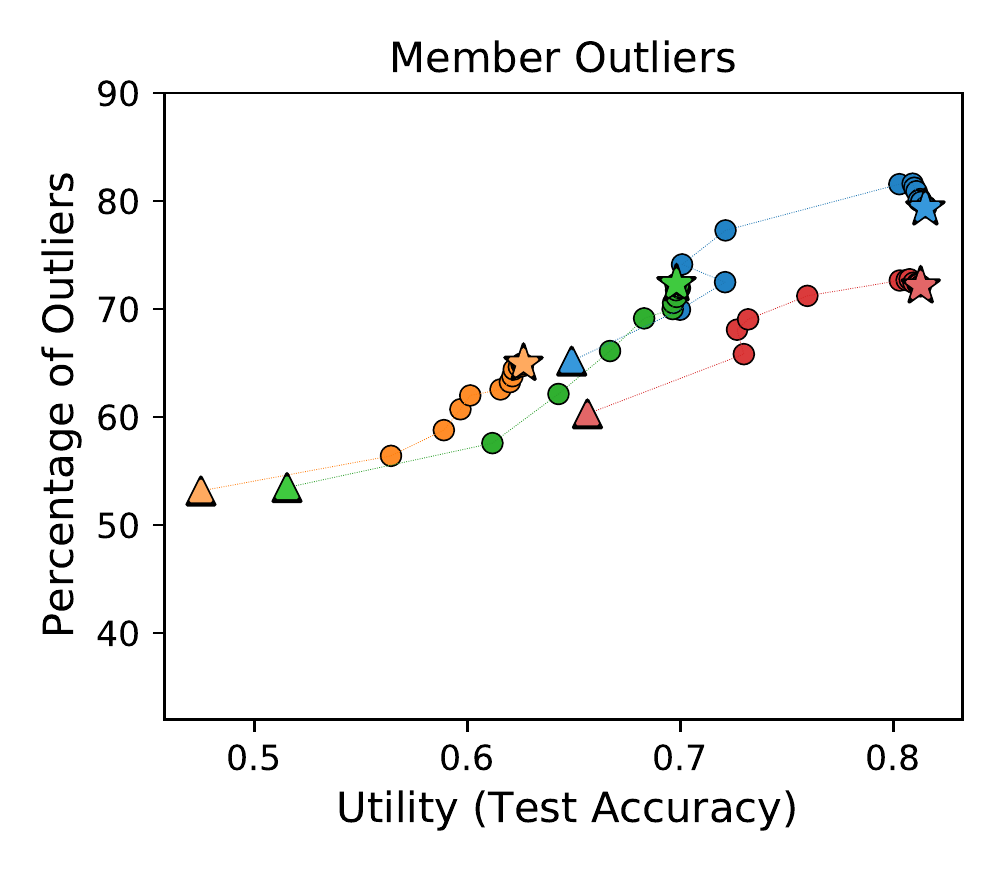}\\
		(a) Member Outliers
	\end{minipage}
	\hfill
	\begin{minipage}{0.32\linewidth}
	\centering
		\includegraphics[width=\linewidth]{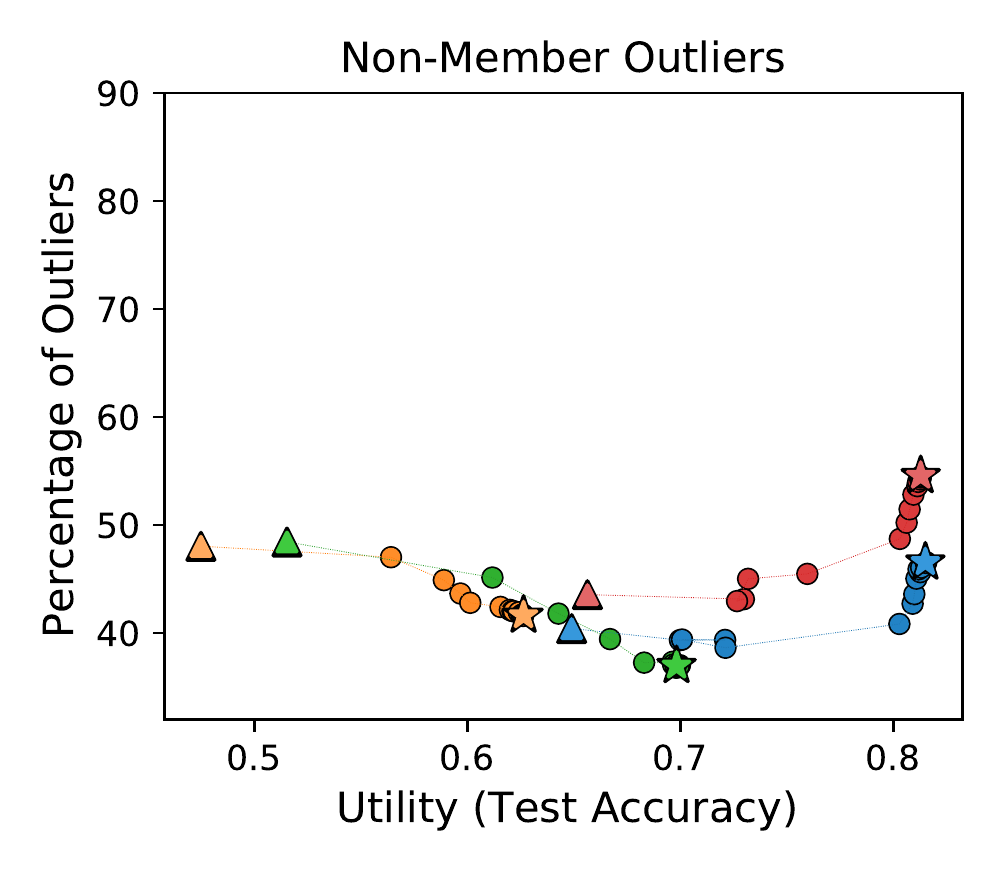}
		(b) Non-Member Outliers
	\end{minipage}
	\hfill
		\begin{minipage}{0.32\linewidth}
	\centering
		\includegraphics[width=\linewidth]{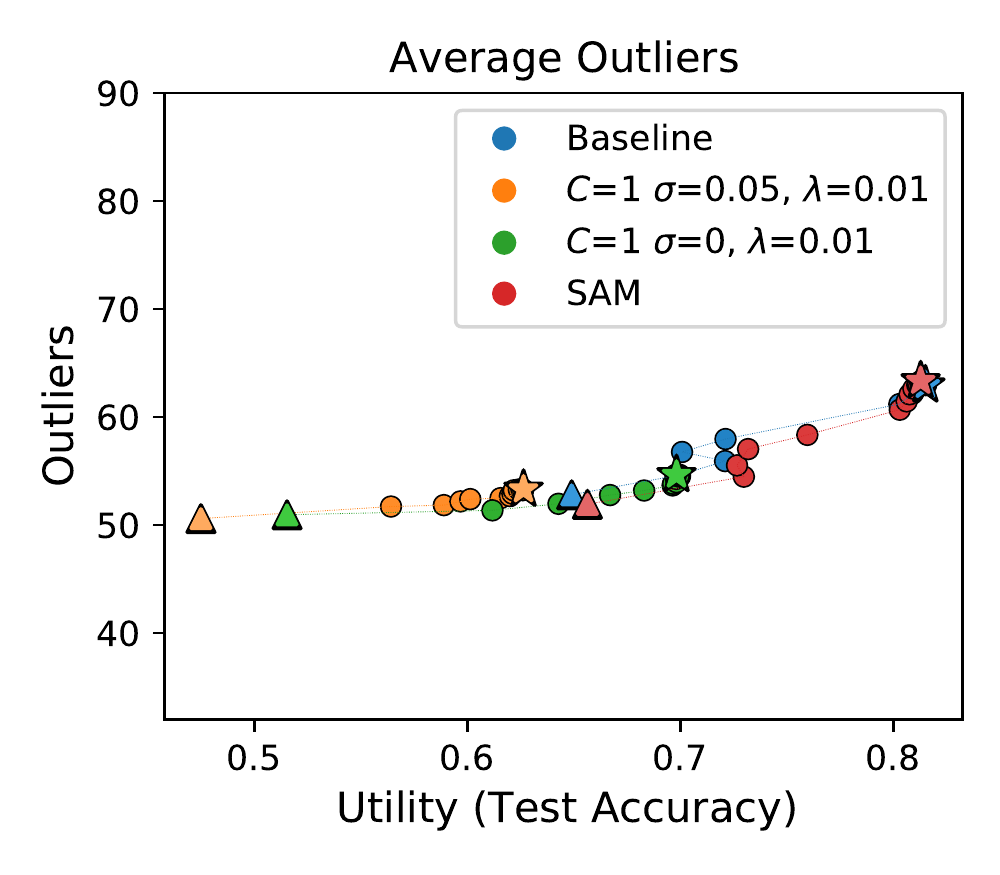}\\
		(c) Average Outliers
	\end{minipage}
	\caption{Percentage of outliers against the threshold attack (i.e., samples the attack identified in 10 out of 10 runs).}\label{fig:exp_out}
\end{figure*}

We identify outliers in our CIFAR-100 experiments for the baseline SGD, for DP-SGD with $\sigma=0$ and $\sigma=0.05$, and for SAM.
We show the percentage of member and non-member outliers vs.~the model's accuracy in Figures~\ref{fig:exp_out}(a) and \ref{fig:exp_out}(b), respectively.
Figure~\ref{fig:exp_out}(c) shows the average of (a,b).
Interestingly, in terms of average number of outliers (Fig.~\ref{fig:exp_out}(c)), all the mechanisms we show approximately lie in the same curve.
This means that, by choosing an appropriate epoch to tune the utility, the percentage of outliers in any of these mechanisms is very similar.
However, when we look at the performance separately over members and non-members, we see that these mechanisms do not protect them evenly.
Overall, we see that SAM provides more protection than other mechanisms for member outliers, and leaves non-member outliers more exposed.
The opposite is true of DP-SGD: we measured a higher number of member outliers when using this technique than when using SAM, and a lower number of non-member outliers.

The fact that DP-SGD protects non-member outliers more than member outliers is a serious privacy issue: typically, correctly identifying the fact that a sample belongs to the training set is more privacy-sensitive than identifying the fact that a non-member was not in the training set.
SAM protects members more than non-members (compared to DP-SGD), which is a more desirable situation from a privacy standpoint.

\subsection{Simple Classification Problem}
For the following experiments, we consider CIFAR-10 dataset, which represents a simpler classification problem.
We perform an initial grid search for DP-SGD using the clipping thresholds $C\in\{0.5, 1, 10\}$ and learning rates $\lrate\in\{0.0005, 0.001, 0.002\}$.
We choose a relatively small $\epsilon=3$, which we achieve by setting the noise multiplier to $\sigma= 0.651$.
We find that $C=1$ and $\lrate=0.001$ performs best, achieving $\perror\approx 0.5$ and test accuracy $\approx0.86$.
This experiment also confirms that lower $\epsilon$ values are feasible in simpler classification problems.

We then compare this configuration of DP-SGD with the baseline SGD, SAM, SAM with $C=1$, and DP-SAM with $C=1$ and $\epsilon=3$ ($\sigma=0.651$).
We show the results in Figure~\ref{fig:exp42_yeom}.
Comparing with our results in CIFAR-100 (Fig.~\ref{fig:exp31}), we see that we can achieve higher privacy and utility levels in CIFAR-10, since the classification problem is easier.
Qualitatively, our conclusions are the same in both datasets: by looking at the performance across all epochs, we see that SAM (with clipping for lower utility values) outperforms both DP-SGD and DP-SAM.

%% file: relwork.tex
\section{Related Work}
It is well-known that generalization techniques protect against MIAs, which explains why many related works evaluate MIA performance against $\ell_2$-regularization, dropouts, and other techniques~\cite{shokri2016membership, salem2018ml, nasr2018machine, choquette2021label, li2020membership}.
Recent works started to consider pre-trained models as well~\cite{choquette2021label, song2021systematic}, and, to the best of our knowledge, we are the first to test the SAM optimizer against membership inference.
However, these works mostly judge the performance of the model at the last epoch, which we have seen can be misleading.
The exception to this, and perhaps the closest work to us, is the work by \citet{song2021systematic}, where the authors show that early stopping (without any additional defense) provides no worse protection than adversarial regularization~\cite{nasr2018machine}, one dedicated defense against MIAs.
In this work we studied the privacy-utility performance of DP-SGD after every training epoch and found that the non-DP variant ($\sigma=0$) performs best than the privacy counterparts, which adds to the findings of \citet{song2021systematic}.
We additionally found that SAM further improves over DP-SGD and the baseline SGD.
Finally, inspired by the work of \citet{long2018understanding} and \citet{choquette2021label}, we studied the protection that DP-SGD offers against particularly vulnerable samples (outliers), and found that SAM is more efficient than DP-SGD towards protecting member outliers.

%% file: conclusion.tex
\section{Conclusions and Future Work}
In this work, we have shown the limitations of Differential Privacy (DP) as an empirical privacy defense against Membership Inference Attacks (MIAs).
We have evaluated the privacy and utility performance of different techniques, including standard SGD, a state-of-the-art DP-SGD implementation, and the recent Sharpness-Aware Minimization optimizer.
By evaluating the performance of the models at each training epoch, we have seen that early stopping, without using any noise, provides a better privacy-utility performance than mechanisms that use DP noise.
We have also seen that DP leaves certain (outlier) members particularly vulnerable against MIAs.
Out of the generalization techniques we tested, SAM seems particularly promising.

This work illustrates the limitations of DP against MIAs, but we believe we have only scratched the surface of this issue.
We have seen that early stopping is critical towards providing a good privacy-utility performance, but more research on how to choose the privacy-utility point needs to be done, perhaps by trying different adaptive learning rate mechanisms or clipping thresholds.


%% file: appendix.tex
\appendices
\section{Additional Experiment Results}
\label{app:results}

\begin{figure*}[t]
	\begin{minipage}{\linewidth}
	\centering
	\begin{minipage}{0.4\linewidth}
	    \centering
		\includegraphics[width=\linewidth]{img/exp0_yeom.pdf}\\
		(a) Threshold Attack
	\end{minipage}\hspace{2cm}
	\begin{minipage}{0.4\linewidth}
	    \centering
		\includegraphics[width=\linewidth]{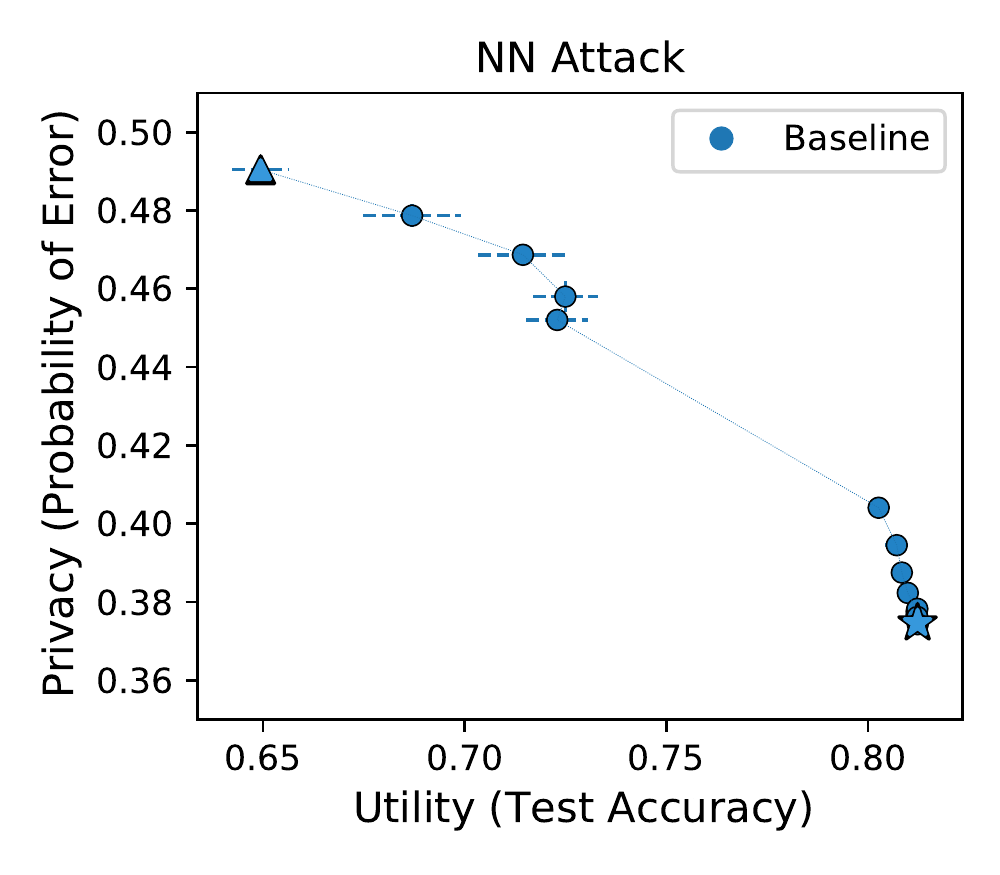}\\
		(b) NN Attack
	\end{minipage}
	\caption{Baseline experiment (CIFAR-100).}\label{fig:exp0}
	\end{minipage}
\end{figure*}

\begin{figure*}[t]
	\begin{minipage}{\linewidth}
	\centering
	\begin{minipage}{0.4\linewidth}
	    \centering
		\includegraphics[width=\linewidth]{img/exp21_yeom.pdf}\\
		(a) Threshold Attack
	\end{minipage}\hspace{2cm}
	\begin{minipage}{0.4\linewidth}
	    \centering
		\includegraphics[width=\linewidth]{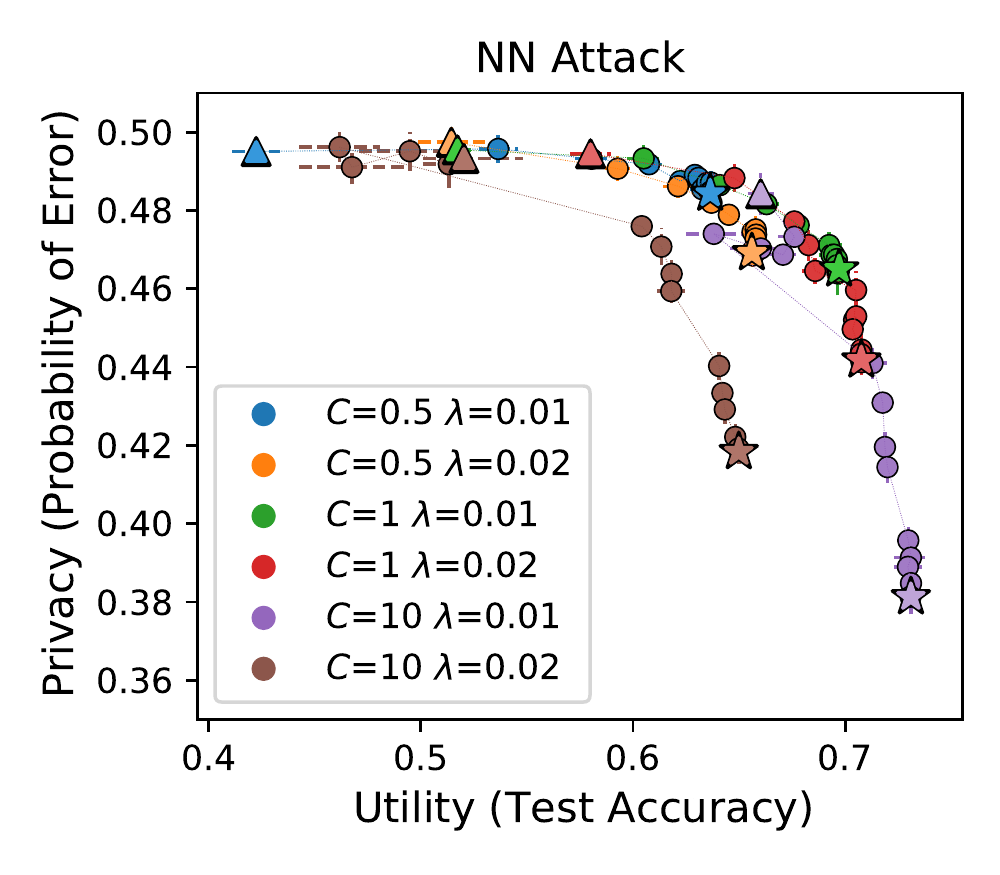}\\
		(b) NN Attack
	\end{minipage}
	\caption{DP-SGD with different clipping thresholds $C$ and learning rates $\lrate$, with $\sigma=0.01$  (CIFAR-100).}\label{fig:exp21}
	\end{minipage}
\end{figure*}

\begin{figure*}[t]
	\begin{minipage}{\linewidth}
	\centering
	\begin{minipage}{0.4\linewidth}
	    \centering
		\includegraphics[width=\linewidth]{img/exp22_yeom.pdf}\\
		(a) Threshold Attack
	\end{minipage}\hspace{2cm}
	\begin{minipage}{0.4\linewidth}
	    \centering
		\includegraphics[width=\linewidth]{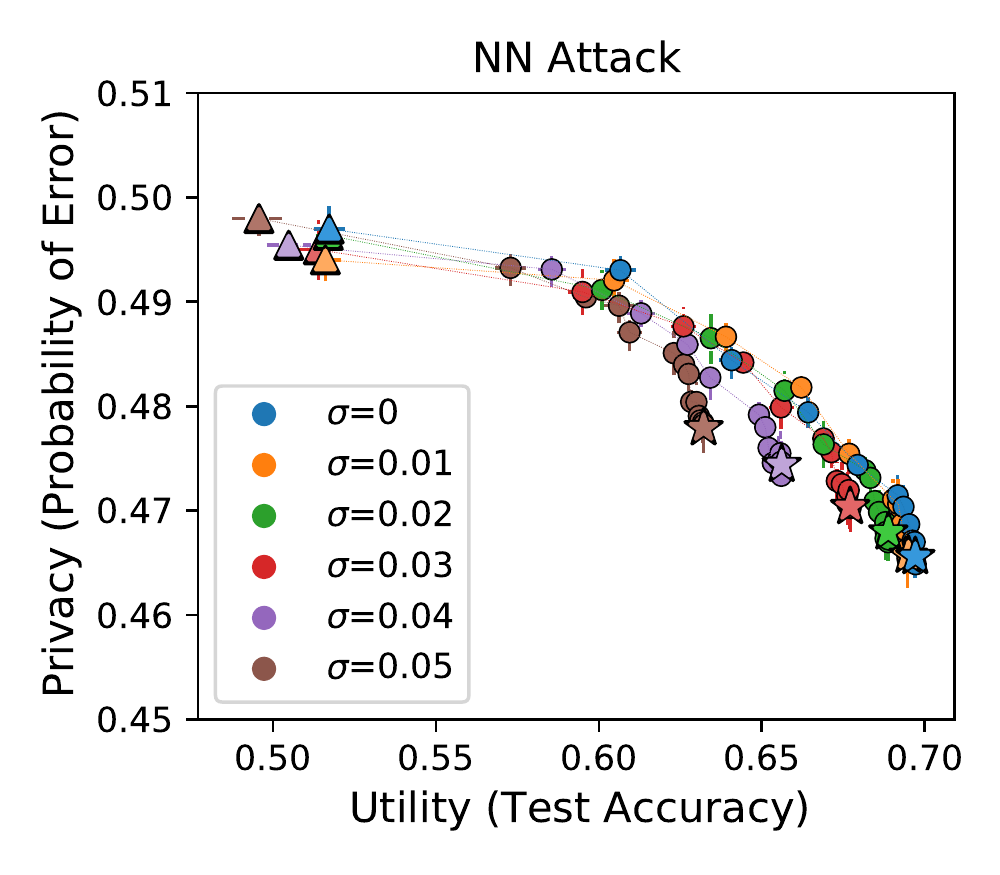}\\
		(b) NN Attack
	\end{minipage}
	\caption{DP-SGD with different noise magnitudes $\sigma$, with $C=1$ and $\lrate=0.01$  (CIFAR-100).}\label{fig:exp22}
	\end{minipage}
\end{figure*}

\begin{figure*}[t]
	\begin{minipage}{\linewidth}
	\centering
	\begin{minipage}{0.4\linewidth}
	    \centering
		\includegraphics[width=\linewidth]{img/exp1_yeom.pdf}\\
		(a) Threshold Attack
	\end{minipage}\hspace{2cm}
	\begin{minipage}{0.4\linewidth}
	    \centering
		\includegraphics[width=\linewidth]{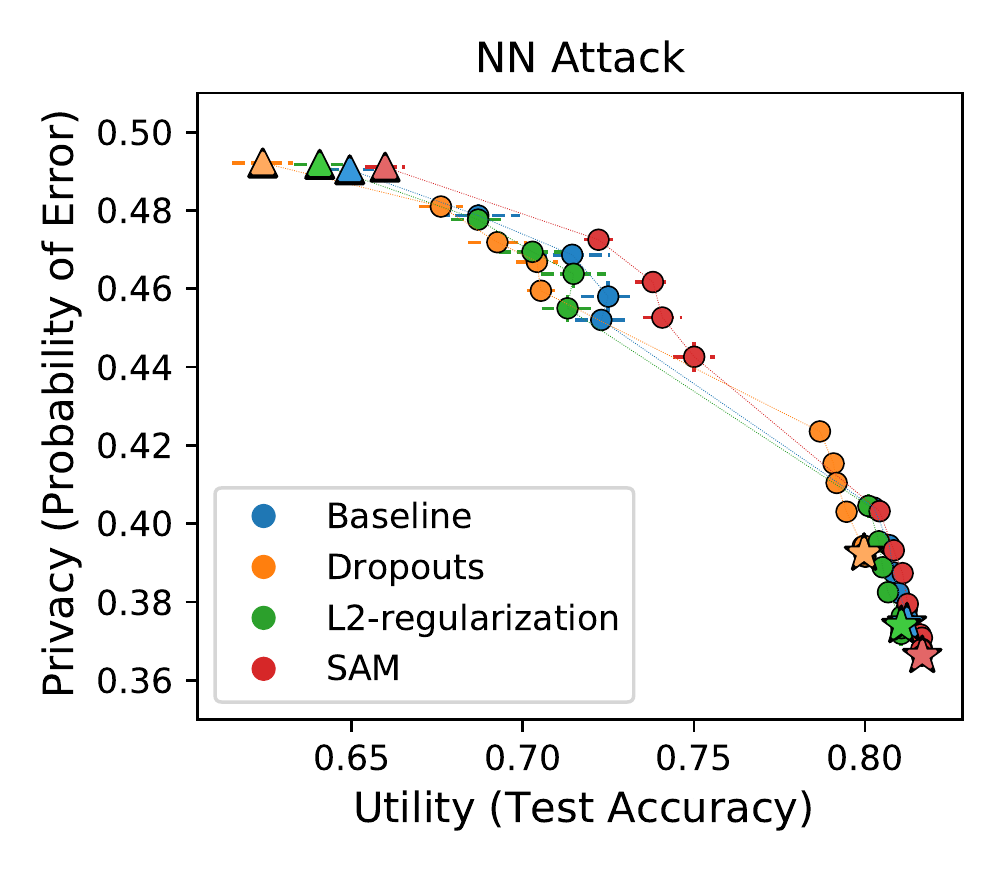}\\
		(b) NN Attack
	\end{minipage}
	\caption{Privacy-utility trade-off for different regularization techniques (CIFAR-100).}\label{fig:exp1}
	\end{minipage}
\end{figure*}

\begin{figure*}[t]
	\begin{minipage}{\linewidth}
	\centering
	\begin{minipage}{0.4\linewidth}
	    \centering
		\includegraphics[width=\linewidth]{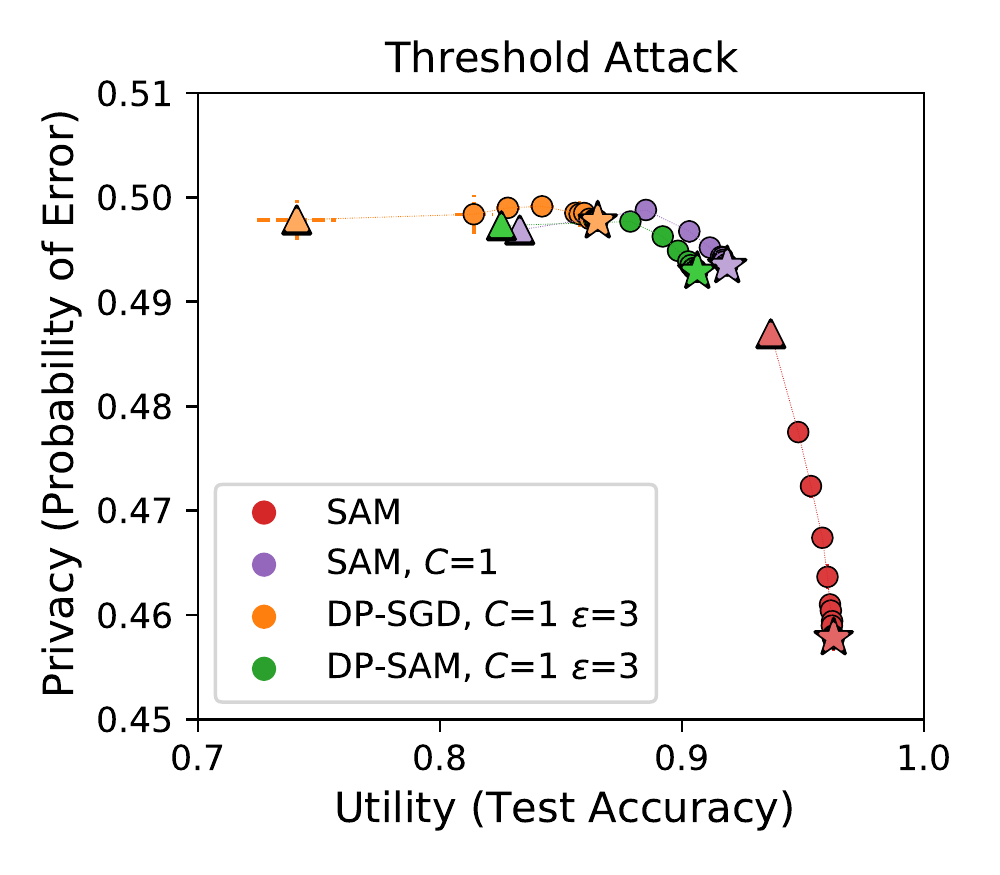}\\
		(a) Threshold Attack
	\end{minipage}\hspace{2cm}
	\begin{minipage}{0.4\linewidth}
	    \centering
		\includegraphics[width=\linewidth]{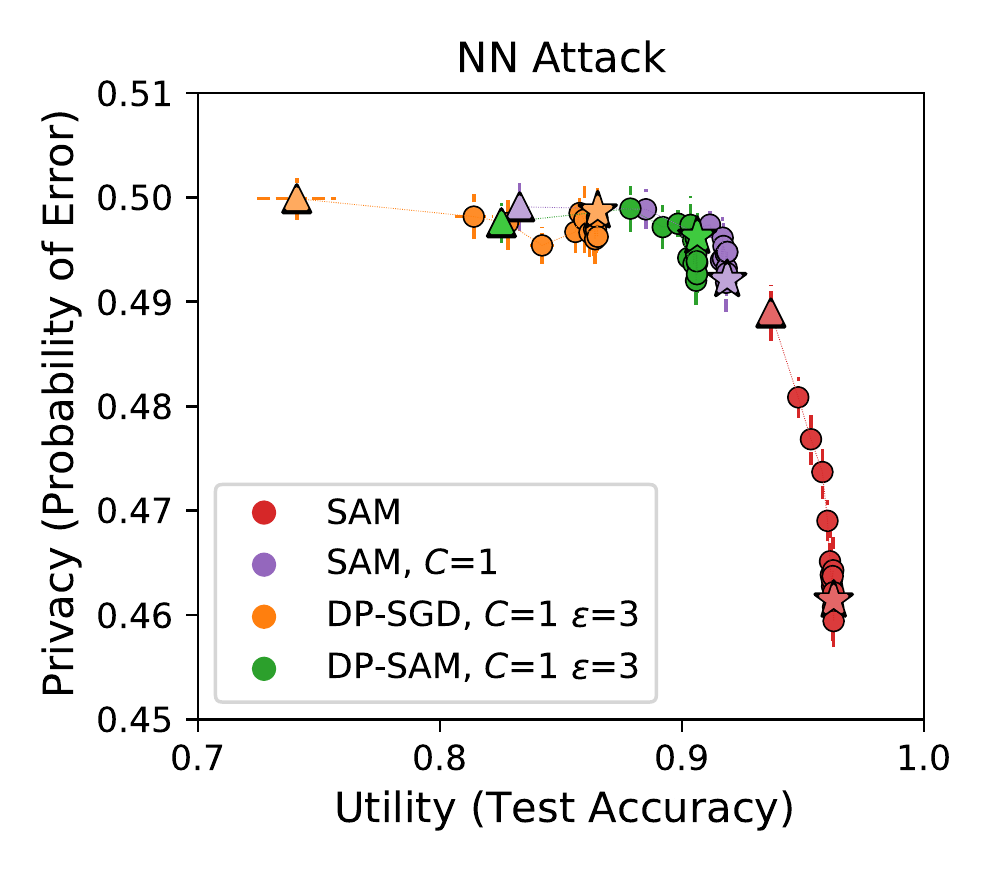}\\
		(b) NN Attack
	\end{minipage}
	\caption{Comparison between SAM, SAM with clipping, DP-SGD, and DP-SAM in CIFAR-10.}\label{fig:exp42}
	\end{minipage}
\end{figure*}

In the main text, we include the results of all our experiments for the threshold attack~\cite{yeom2017privacy}, and we point out that the results for the NN attack~\cite{salem2018ml} are similar and do not add anything to our findings.
In order to support this claim, we provide a side-by-side comparison of the results for the threshold and the NN attacks in Figures~\ref{fig:exp0}--\ref{fig:exp42}.

\begin{figure*}[t]
	\begin{minipage}{\linewidth}
	\centering
	\begin{minipage}{0.4\linewidth}
	    \centering
		\includegraphics[width=\linewidth]{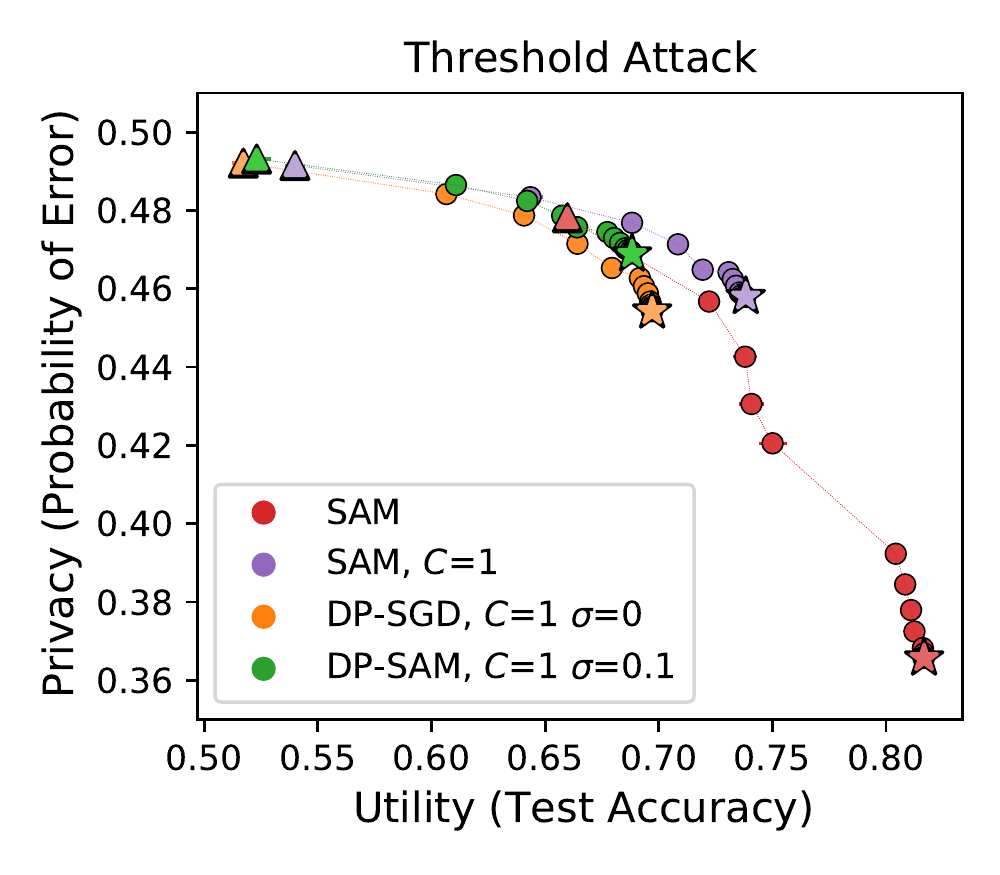}\\
		(a) Threshold Attack
	\end{minipage}\hspace{2cm}
	\begin{minipage}{0.4\linewidth}
	    \centering
		\includegraphics[width=\linewidth]{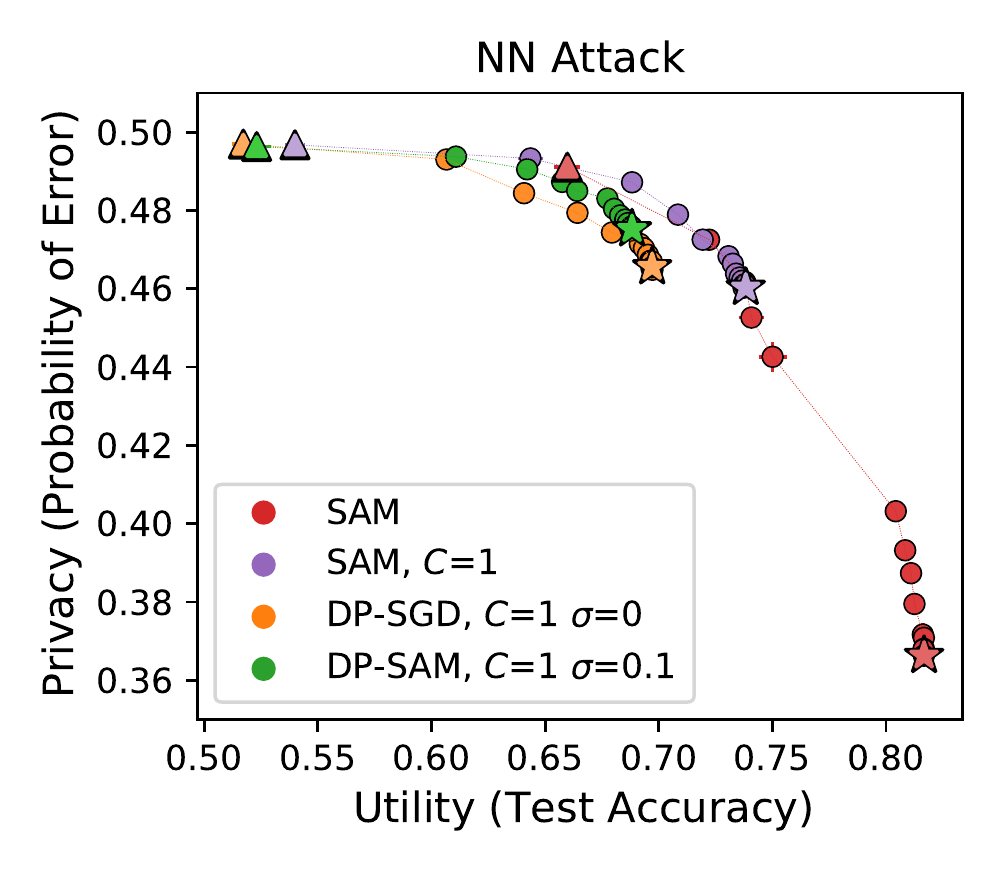}\\
		(b) NN Attack
	\end{minipage}
	\caption{Comparison between SAM, SAM with clipping, DP-SGD, and DP-SAM in CIFAR-100.}
	\end{minipage}
\end{figure*}

\begin{figure*}[t]
	\begin{minipage}{\linewidth}
	\centering
	\begin{minipage}{0.4\linewidth}
	    \centering
		\includegraphics[width=\linewidth]{img/exp42_yeom.pdf}\\
		(a) Threshold Attack
	\end{minipage}\hspace{2cm}
	\begin{minipage}{0.4\linewidth}
	    \centering
		\includegraphics[width=\linewidth]{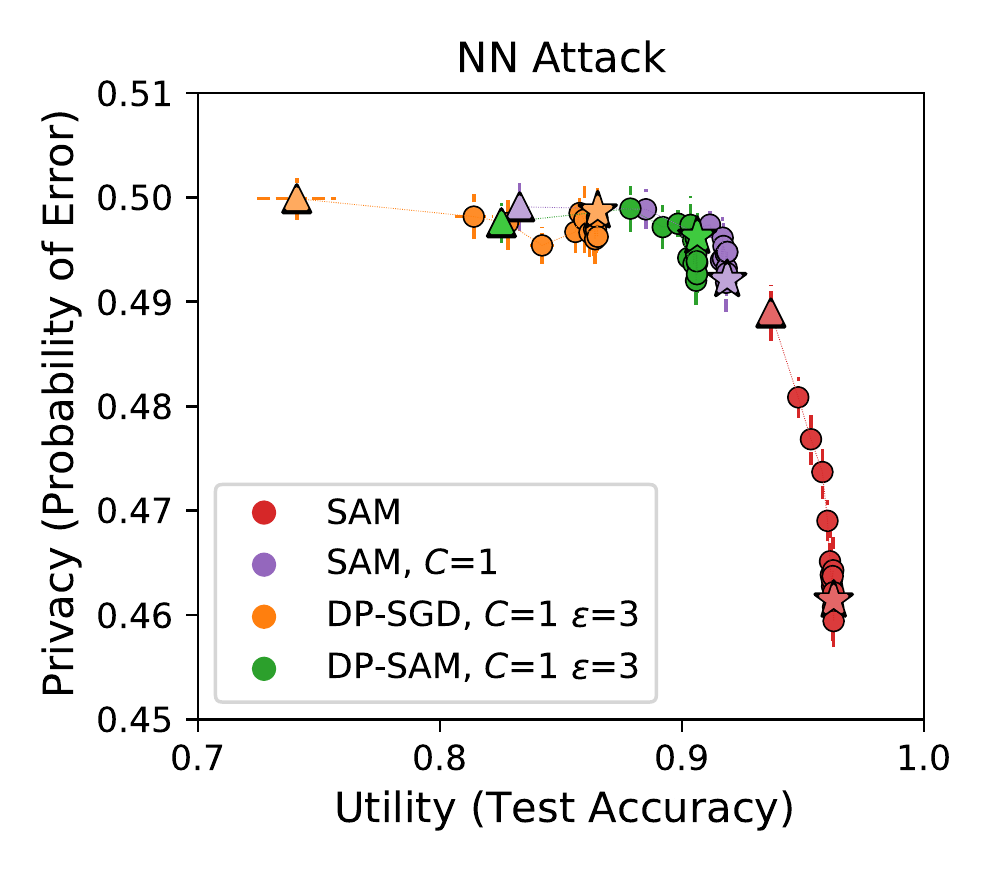}\\
		(b) NN Attack
	\end{minipage}
	\caption{Comparison between SAM, SAM with clipping, DP-SGD, and DP-SAM in CIFAR-10.}
	\end{minipage}
\end{figure*}

\section{Pseudo-code for DP-SAM}
\label{app:DP-SAM}

We implement DP-SAM by plugging Opacus' Privacy Engine into the SAM implementation.
We summarize here the resulting algorithm.
Let $\wei$ be the model's weights, and let $\loss(\wei)$ be the training set loss.
SAM~\cite{foret2020sharpness} re-defines the loss function as
\begin{equation} \label{eq:SAM}
    \loss^{SAM}(\wei)\doteq \max_{||\eee||_2\leq \rho} \loss(\wei+\eee)\,.
\end{equation}
This means that minimizing $\loss^{SAM}(\wei)$ requires looking for weights $\wei$ such that all neighbouring weights (at a certain $\ell_2$-distance around $\wei$) also yield low loss (i.e., a wide minimum).
\citeauthor{foret2020sharpness} provide details on how to implement this minimization efficiently, which includes a formula on how to compute $\eee$ given $\wei$.
We refer to the original paper for details~\cite{foret2020sharpness}, and here we just use $\hat{\mathbf{g}}_i(\wei)$ to denote the gradient approximation of the SAM objective function \eqref{eq:SAM} for the $i$th sample in the batch $z_i\in\mathcal{B}$ .

We represent DP-SAM in Algorithm~\ref{alg}.
DP-SAM first uses the batch samples to estimate the gradient that minimizes the SAM objective function.
Then, it clips the gradients evaluated at each sample in the batch, averages them, and adds Gaussian noise.
The clipping plus noising operation ensures differential privacy following DP-SGD~\cite{abadi2016deep}.

\begin{algorithm}
\caption{DP-SAM}\label{alg}
\begin{algorithmic}[1]
\REQUIRE Training set $S\in\mathcal{Z}^n$, batch size $b$, step size $\lrate>0$, neighbourhood size $\rho>0$, clipping threshold $C>0$, noise multiplier $\sigma>0$.
\ENSURE Trained model
 \STATE Initialize model's weights $\mathbf{w}_0$, $t=0$.
 \WHILE{not converged} 
    \STATE Sample batch $\mathcal{B}=\{(x_1,y_1),\dots,(x_b,y_b)\}$;
    \FOR{each $z_i\in\mathcal{B}$}
    \STATE Compute the SAM gradient: $\mathbf{g}_i=\hat{\mathbf{g}}_i(\wei)$;
    \IF{$||\mathbf{g}_i||_2>C$} \STATE{Clipping: $\mathbf{g}_i=\mathbf{g}_i\cdot C/||\mathbf{g}||_2$;} \ENDIF
    \ENDFOR
    \STATE Add Gaussian noise $\mathbf{g}=\frac{1}{b}\sum_{i=1}^b\mathbf{g}_i+\mathcal{N}(0, \sigma^2 C^2\mathbf{I})$;
    \STATE Update weights $\wei_{t+1}=\wei_t-\lrate\mathbf{g}$;
    \STATE $t=t+1$;
\ENDWHILE
\end{algorithmic}
\end{algorithm}